\documentclass[epjCONF]{svjour}
\usepackage{graphicx}
\usepackage[varg]{txfonts} 
\usepackage[latin1]{inputenc}
\session-title{Hot and Cold Baryonic Matter -- HCBM 2010}

\newcommand{\be}{\begin{equation}}
\newcommand{\ee}[1]{\label{#1} \end{equation}}
\newcommand{\ba}{\begin{eqnarray}}
\newcommand{\ea}[1]{\label{#1} \end{eqnarray}}
\newcommand{\nl}{\nonumber \\}

\begin{document}
\title{Pion Production Via Resonance Decay in a Non-extensive Quark-Gluon Medium with Non-additive Energy Composition Rule
}
\author{K. Urmossy\inst{1,2}\fnmsep\thanks{\email{karoly.uermoessy@cern.ch}} \and T. S. Biro\inst{2} \and G. G. Barnaf\"oldi\inst{2} }
\institute{Department for Theoretical Physics, E\"otv\"os University, 1/A P\'azm\'any P\'eter s\'et\'any, H-1117 Budapest, Hungary\and Research Institute for Particle and Nuclear Physics of the HAS, 29-33, Konkoly-Thege Mikl\'os Str.,  H-1121 Budapest, Hungary }

\abstract{
Resonance production and decay into pion pairs is simulated in a non-extensive quark matter with multi-particle interactions. Final state pion spectra are found to take the form of the Tsallis distribution, in accordance with measurements. It has also been shown that, if a large number of particles with these multi-particle interactions are constrained to a constant energy hyper-surface in phase space, the one-particle distribution is the Tsallis distribution.
} 
\maketitle
\section{Introduction}
\label{sec:intro}
Transverse hadron spectra measured in high energy collisions in the last three decades, fit to the Tsallis distribution (TS) (see Refs.~\cite{bib:tsinppNN1}-\cite{bib:tsinppNN10} for proton-proton ($pp$), proton-antiproton ($p\bar{p}$) and nucleus-nucleus ($AA$) collisions and Refs.~\cite{bib:tsinee1,bib:tsinee2} for $e^+e^-$ collisions). On the theoretical side, there are many proposals on the emergence of the TS distribution. In kinetic theory, the collision term of the Boltz\-mann-equa\-tion \cite{bib:BEq1}-\cite{bib:BEq3}, or the noise term of the Langevin-equation \cite{bib:noisy1} can be generalised in a way, in which the TS distribution is the stationary sollution.
In equilibrium thermodynamics, the \textit{Maximum Entropy Principle} (MEP) together with a generalisation of the Shannon-entropy formula (the Tsallis-entropy, see Ref.~\cite{bib:next1}) also lead to the TS distribution, as a generalisation of the canonical Boltzmann-Gibbs distribution (BG). The TS distribution can also be derived from the MEP by introducing special interactions, while, leaving the original Shannon-entropy unaltered \cite{bib:tsinppNN10a,bib:comp1}.

In all the above cases, equilibrium, or at least stationarity is assumed, while the question of equilibration in high energy collisions is still a subject of intense debate. However, examining the mathematical foundations of statistical physics \cite{bib:math1,bib:math2}, it turns out, that statistical physical distributions may be used not only in equilibrium. Whenever a system is composed of idependent and identically distributed particles, and the total energy of the system is conserved, the one-particle distribution may be approximated by the canonical distribution in the limit of a large number of particles. The reason for this is that \textit{Central Limit Theorems} (CLT) do not deal with the issue, whether particles have the same distribution, because they are thermalised, or because they are produced via the same process.

In this paper, we discuss resonance production in a non-extensive quark-gluon medium, in which multiparticle interactions of the form of Eq.~(\ref{onTS4}) are present. In Sect.~\ref{sec:onTS} we show, that (apart from a phase-space factor) the one-particle energy distribution in such a medium is the TS distribution in the limit of a large number of particles. In the calculation we use only probability theory, thus the result is valid not only in equilibrium. We also outline some advantages and disadvantages of the probability theory approach to the thermodynamical one.

In Sect.~\ref{sec:pccs} we introduce a model in which a non-ex\-ten\-sive quark matter ($QM$) evolves in time via the collisions of randomly chosen quark pairs. In the collisions, the 3-mo\-men\-tum, and the total (non-extensive) energy of the system is conserved. In Sect.~\ref{sec:Rprod} we present a way, in which hadron resonances may leave the $QM$ and decay into pion pairs without the violation of the conservation of the total energy and momentum of the system.

Sect. \ref{sec:res} contains the resulting final-state pion spectra and the mass spectrum of the resonances. Sect. \ref{sec:con} contains our concluding remarks.

\section{Probability Theory, Maximum Entropy Principle and the Tsallis Distribution}
\label{sec:onTS}
In this section we review the derivation of the canonical BG distribution using probability theory. During the deduction, we allow for a constraint on the N particle phase space that is more general than the conservation of the sum of the one-particle energies. This way, we may obtain a wider class of distributions covering those having power-law asymptotics as well.

Let us assme, that a system is composed of a large number of particles, all being independent and identically distributed (iid.). For example let the particles have the same phase space structure with momentum-space distribution, $dF(\vec{p_i})$ (in a homogenious, isotropic ensemble \\$dF(\vec{p_i})~\propto~d^3 \vec{p_i}$). In this case the probability that a particle has energy, $\epsilon$, while the total energy of the system, $E$ is fixed, is

\be
\frac{d\mathcal{P}_N}{dF(\vec{p})} = \frac{\Omega_{N-1}(E')}{\Omega_{N}(E)}
\ee{onTS1}
with $E'$ being the energy of $N-1$ particles (when particles do not interact, $E'=E-\epsilon$) and $\Omega_{N}(E)$ being the phase space volume of $N$ particles restricted to the constant energy hyper-surface:

\be
\Omega_{N}(E) = \int \prod _i dF(\vec{p_i}) \, \delta\left( \sum _j L(\epsilon_j) - L(E) \right)
\ee{onTS2}
via the constraint,

\be
\sum _j L(\epsilon_j) = L(E) \, .
\ee{onTS3}
For example, if the particles are non-interacting, $\sum \epsilon_j = E$, thus $L(\epsilon)=\epsilon$. In the model, reported in \cite{bib:tsinppNN10a}, the following multiparticle interaction has been proposed:

\ba
E_N &=& \epsilon_1 + \dots + \epsilon_N \nl
 &+& a \, ( \, \epsilon_1 \epsilon_2 + \dots + \epsilon_{N-1} \epsilon_N \, ) \nl
&\vdots& \nl &+&  a^{N-1} \epsilon_1 \cdots \epsilon_N.
\ea{onTS4}
The above rather complicated energy formula can be turned into a simlpe addition of the form of Eq.~(\ref{onTS3}) by a so-called \textit{formal logarithm} \cite{bib:comp1},

\be
L(\epsilon) = \frac{1}{a} \ln(1 + a\,\epsilon) \, .
\ee{onTS5}
Moreover, the constrained $N$ particle phase space volume, Eq.~(\ref{onTS2}) can be factorised using $2\pi\,\delta(x) = \int ds\, \exp(i\,s\,x)$:

\be
\Omega_{N}(E) = \int \limits _{-\infty} ^\infty ds\;  \exp \left\{-\,N\,\left( \,-i\,s \frac{L(E)}{N} - \hat{I}(s)  \right) \right\}
\ee{onTS6}
with the logarithmic generator function of the one-particle phase\-space distribution, $\hat{I}(s) = \ln \left( \int dF(\vec{p}) \exp \left[ -i\, s\, L(\epsilon) \right] \right)$. In the large $N$ limit, Eq.~(\ref{onTS6}) may be approximated by the \textit{saddle-point method}:

\be
\Omega_{N}(E) \approx \exp \left\{-\,N\,\mathcal{F}[ s^{\star}] \right\}
\ee{onTS7}
where $\mathcal{F}$ is the free energy per particle,

\be
\mathcal{F}[s^{\star}] = -i\,s^{\star} \frac{L(E)}{N} - \hat{I}(s^{\star})\, ,
\ee{onTS8}
and the inverse temperature $\beta = i\,s^\star$ minimises $\mathcal{F}$:

\be
\frac{L(E)}{N} = \frac{ \int dF(\vec{p}) \, L(\epsilon) \, \exp \left[ -\beta\, L(\epsilon) \right]}
{\int dF(\vec{p}) \exp \left[ -\beta\, L(\epsilon) \right]} \,.
\ee{onTS9}
Note, that $\beta$ and thus $\mathcal{F}$ also depend on $L(E)/N$. Applying Eq.~(\ref{onTS7}), the one-particle energy distribution (Eq.~(\ref{onTS1})) in the large N limit, is  approximately

\be
\frac{d\mathcal{P}_N}{dF(\vec{p})} \approx 
\exp\left\{ -(N-1)\,\mathcal{F}\left[ \frac{L(E')}{N-1} \right] + N\,\mathcal{F}\left[ \frac{L(E)}{N} \right] \right\}
\ee{onTS10}
with $L(E') = L(E) - L(\epsilon)$. Exploiting that $E\gg \epsilon$, in the first term in the bracket, we may Taylor-expand $\mathcal{F}$ around $L(E)/N$. It follows from Eq.~(\ref{onTS8}) that

\be
\mathcal{F}'\left[\frac{L(E)}{N}\right]
= -i\,s^\star + \left( -i\, \frac{L(E)}{N} - \hat{I}'(s^{\star}) \right)\,s^{\star\prime}
= -\beta \, .
\ee{onTS11}
The expression in the bracket in Eq.~(\ref{onTS11}) vanishes because of Eq.~(\ref{onTS9}). Consequently, if we neglect the terms in the Taylor-series that are proportional to $1/N$ (or smaller), the one-particle distribution, Eq.~(\ref{onTS10}) gives the distribution:

\be
\frac{d\mathcal{P}_N}{dF(\vec{p})} \approx \frac{ \exp\{-\beta L(\epsilon) \} }{ \int dF(\vec{p})\, exp \{-\beta L(\epsilon) \}} \,.
\ee{onTS12}

The above calculation is similar to the variational met\-hod based on the \textit{Maximum Entropy Principle}. There, a free energy functional $\Phi[f]$ is constructed from the one-particle distribution $f$, that contains an entropy term, $S[f]$ and constraints $C[f]$:

\be
\beta \,\Phi[f] = -S[f] + \beta\, C[f] \,.
\ee{onTS13}
There are several proposed formulas on $S[f]$, however there is no recipe on how to construct the appropriate $S[f]$ for a system defined by a given Hamiltonian. The following choices of $C[f]$ and $S[f]$ both lead to the TS distribution:

\ba
S[f] &=& -\int dF(\vec{p})\,f(\epsilon) \ln[f(\epsilon)]\nl
C[f] &=&  \int dF(\vec{p})\, L(\epsilon)\, f(\epsilon) - \frac{L(E)}{N}  \,;
\ea{onTS14}

\ba
S[f] &=& \frac{1 - \int dF(\vec{p})\,f^{\,q}(\epsilon) }{q-1} \nl
C[f] &=& \int dF(\vec{p})\, \epsilon\, f(\epsilon) - \frac{E}{N}  \,.
\ea{onTS15}
In Eq.~(\ref{onTS14}), the entropy is additive for independent particles, while interactions of the form of Eq.~(\ref{onTS3}) are pressumed, thus the mean one-particle energy is constrained through the formal logarithm.\\
In Eq.~(\ref{onTS15}), particles are assumed to be non-interacting, thus the mean energy per particle is set to be $E/N$, however, a generalised entropy formula accounts for correlations. 

Both the \textit{Maximum Entropy Principle} and the \textit{probability theoretical} approach are based on finding the minimum of a free energy, however, there are two main differences between them:
\begin{description}
\item[1)] Behind the Maximum Entropy Principle, there is the assumption of \textit{thermal equilibrium}, while in the probability theoretical approach, there is \textit{no need for such an assumption};

\item[2)] In the probability theoretical approach, particles \textit{need to be independent}, while this is not required in a model based on the Maximum Entropy Principle.
\end{description}

\section{Parton Collision Cascade Simulation and Pion Production}
\label{sec:pccs}
In this section, we use a model reported in \cite{bib:tsinppNN10a}. We consider $N$ massless quarks ($\epsilon_i = |\vec{p_i}|$) with interactions of the form of Eq.~(\ref{onTS4}). The ensemble has an initial momentum distribution that is homogenious inside a Fermi sphere of radius $p_F$. The momentum distribution evolves in time via collisions of randomly chosen pairs. In pair collisions the 3-momentum and the formal logarithm of the energy are conserved:

\begin{figure}
\begin{center}
 \includegraphics[width=0.5\textwidth]{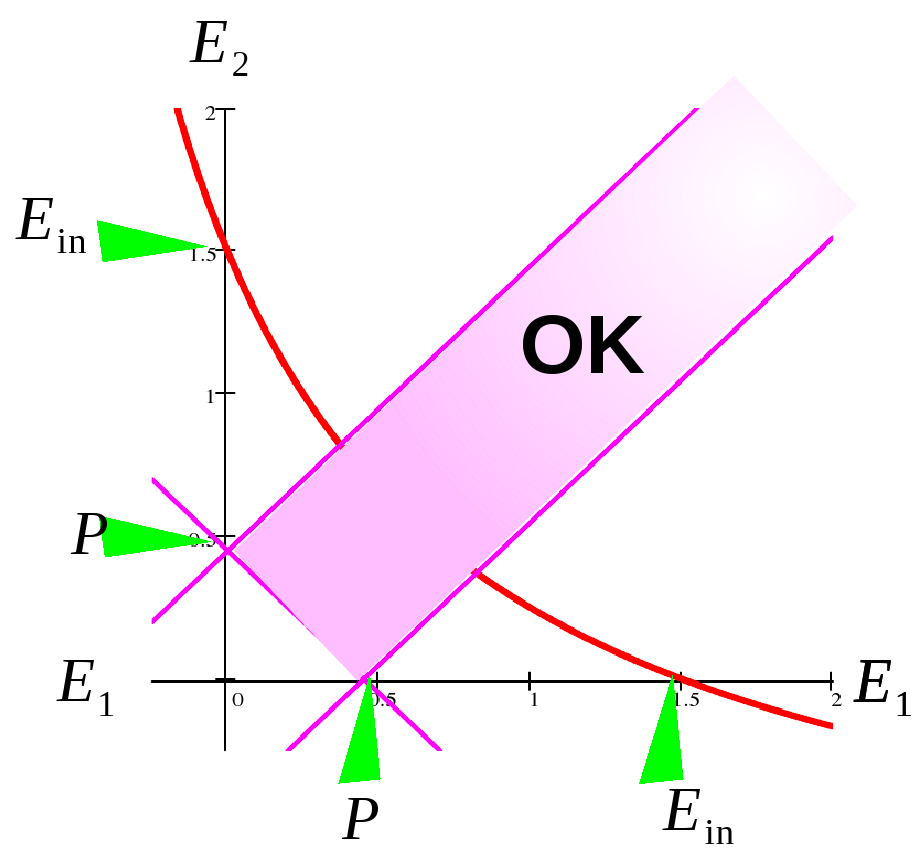}
\end{center}
\caption{The $E_1-E_2$ plane, in case, when the total momentum of the incoming quarks $P$ is sufficiantly small. Because of the triangle inequality among $\vec{p_1}$, $\vec{p_2}$ and $\vec{P}$, the energies of the outgoing quarks ($E_1$ and $E_2$) have to be chosen from the filled area. Furthermore, $E_1$ and $E_2$ have to satisfy Eq.~(\ref{pccs4}) (solid red curve).}
 \label{fig:regions1}
\end{figure}

\ba
\vec{p_1} + \vec{p_2} &=& \vec{p_3} + \vec{p_4}\, ,\nl
L(\epsilon_1) + L(\epsilon_2) &=& L(\epsilon_3) + L(\epsilon_4) \, .
\ea{pccs1}
(The second row in Eq.~(\ref{pccs1}) also ensures the conservation of not just the pair energies but that of the total system.) Thus the momenta of the outgoing quarks are chosen randomly according to the distribution

\ba
dw &=& d^3p_1\, d^3p_2\, \delta\left(\vec{p_1} + \vec{p_2} - \vec{P}\right)  \nl && \times \;\delta(E_1 + E_2 + aE_1E_2 - E_{in} )
\ea{pccs2}
with the total momenta and energy of the incoming quarks: $\vec{P}$ and $E_{in}$. Integrating out for $\vec{p_2}$, we obtain the  energy distribution of the first outgoing quark:

\be
p(E_1) \sim \frac{ E_1\, (E_{in} - E_1) }{(1+aE_1)^2}
\ee{pccs3}
Once $E_1$ is obtained, the energy of the second outgoing quark is

\be
E_2 = \frac{E_{in}-E_1}{1+aE_1}\, .
\ee{pccs4}
The particles are massless, thus the equation $\vec{p_1}+\vec{p_2}=\vec{P}$ defines the angles between their momenta and $\vec{P}$. However, $E_1$ may not take any value between 0 and $E_{in}$, because $\vec{p_1}$, $\vec{p_2}$ and $\vec{P}$ must obey the triangle inequality.

\begin{figure}
\begin{center}
 \includegraphics[width=0.5\textwidth]{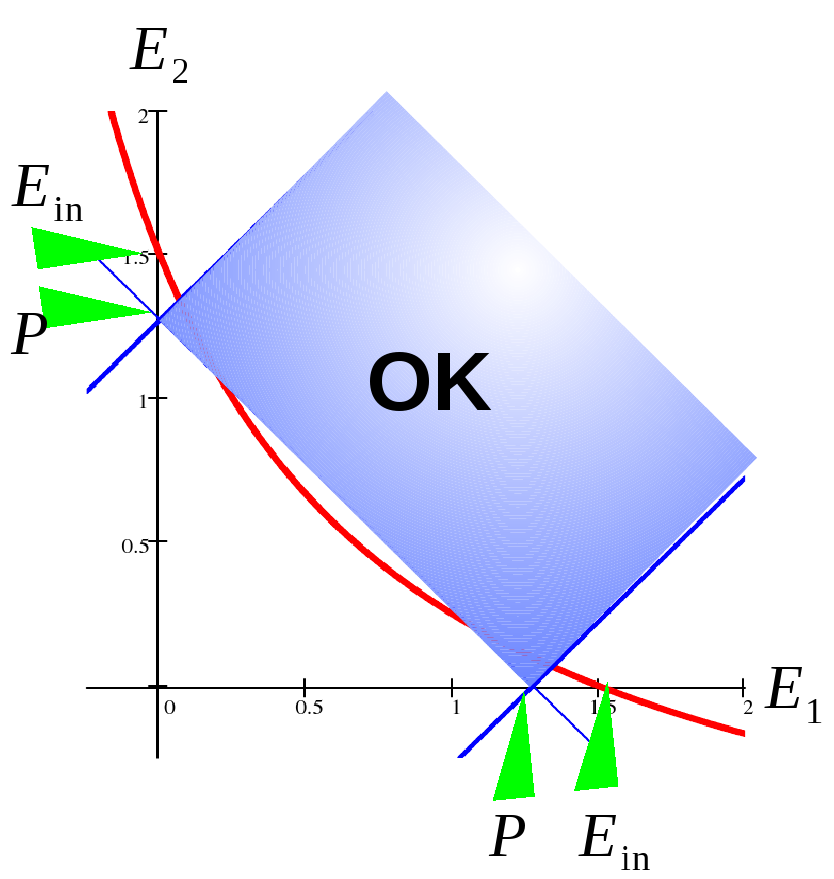}
\end{center}
\caption{The $E_1-E_2$ plane, in case, when the total momentum of the incoming quarks $P$ is sufficiantly large. Because of the triangle inequality among $\vec{p_1}$, $\vec{p_2}$ and $\vec{P}$, the energies of the outgoing quarks ($E_1$ and $E_2$) have to be chosen from the filled area. Furthermore, $E_1$ and $E_2$ have to satisfy Eq.~(\ref{pccs4}) (solid red curve).}
 \label{fig:regions2}
\end{figure}

\begin{description}
\item[$\bullet$] If $P \leq \frac{2}{a}\left(\sqrt{1+a\,E_{in}} -1 \right)\, ,$ we have to choose $E_1$ from the interval $E_1 \in \pm\frac{P}{a} - \frac{1}{a} + \sqrt{ \left(\frac{P}{2}\right)^2 + \frac{1}{a^2} + \frac{E_{in}}{a} }$ (Fig.~\ref{fig:regions1});

\item[$\bullet$] If $P > \frac{2}{a}\left(\sqrt{1+a\,E_{in}} -1 \right)\, ,$ $E_1$ must satisfy the conditions: $E_1 \in \pm\frac{P}{a} - \frac{1}{a} + \sqrt{ \left(\frac{P}{2}\right)^2 + \frac{1}{a^2} + \frac{E_{in}}{a} }$ , and that
$E_1 \notin \frac{P}{2} \pm  \sqrt{ \left(\frac{P}{2}\right)^2 - \frac{E_{in}-P}{a} }$ (Fig.~\ref{fig:regions2}).
\end{description}

An interesting feature of this model is its exponential non-extensivity. Let us estimate the dependence of the total energy of the system on the number of its constituents in the case, when the mean energy per particle, $\bar{\epsilon}$ is fixed. From Eq.~(\ref{onTS3}), the total enrgy of the system estimated from the equipartition principle, $\epsilon_i = \bar{\epsilon}$, is

\be
E \approx L^{-1}[N\,L(\bar{\epsilon})] \, \approx\, \frac{1}{a}(1 + a \bar{\epsilon})^N .
\ee{pccs5}

\subsection{Resonance Production in a Non-extensive Medium}
\label{sec:Rprod}
\begin{figure}
\begin{center}
 \includegraphics[width=0.5\textwidth]{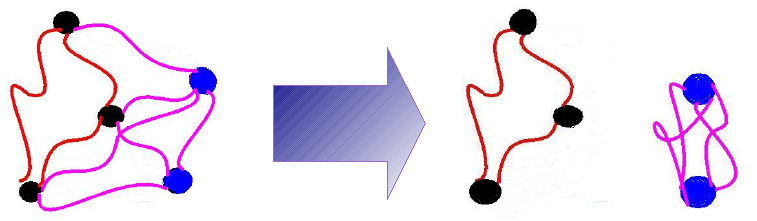}
\end{center}
\caption{Schematic picture of resonance production in a non-extensive medium. Spots represent quarks, curves represent interactions among them.}
 \label{fig:reso1}
\end{figure}

Eq.~({\ref{pccs5}}) illustrates that, when a color-neutral quark anti-quark pair leaves the system, and forms a hadron resonance, it carries a big amount of interaction energy. (see Fig.~\ref{fig:reso1} as a sche\-ma\-tic picture of resonance production) Because of the 3-momentum conservation, this hadron can not be on-shell. For simplicity, we do not distinguish between quark flavours, and consider only a pion pair production in the decay of the resonance.
Since the resonance ($R$) has left the quark matter ($QM$), $R$ does not interact with the $QM$ any more. Thus the energy of $R$ is simply the difference of the total energy of the $QM$ before ($E_{N}$) and after ($E_{N-2}$) the formation of $R$:

\ba
E_R &=& E_{N} - E_{N-2} = E_{12} \frac{1+aE_N}{1+aE_{12}}\nl
&\approx&\, (1+a\bar{\epsilon})^{N-2}(2\bar{\epsilon} + a\bar{\epsilon}^2)
\ea{pccs6}
with $E_{12} = E_1 + E_2 + aE_1E_2$ being the energy of the $q\bar{q}$ pair without its interaction with the rest of the $QM$. The second equality and the last term in Eq.~(\ref{pccs6}) follow from Eq.~(\ref{onTS3}). This shows, that the energy of $R$ can be much greater than that of the $q\bar{q}$ pair, while the momentum of $R$ is exactly the same of that of the $q\bar{q}$ pair. Consequently, the production of big resonance masses is allowed, as shown in Fig.~\ref{fig:dNperdM}. The rest mass of $R$ then contributes to the kinetic energies of the pions when $R$ decays, resulting in a long tail of the momentum distribution of the pions, as shown in Figs.~\ref{fig:dNperdPa1}-\ref{fig:dNperdPaLog}.

\section{Results}
\label{sec:res}

\begin{figure}
\begin{center}
 \includegraphics[width=0.5\textwidth]{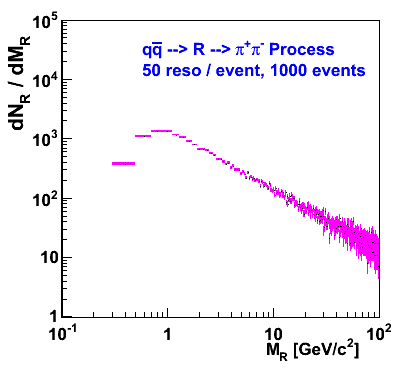}
\end{center}
\caption{Simulated resonance mass distribution from the model described in Sect.~\ref{sec:pccs}. The histogram in the figure is the sum of statistics collected in 1000 events with 50 resonance produced in each event. Through out the simulation, $a=1$ was used. }
 \label{fig:dNperdM}
\end{figure}

\begin{figure}
\begin{center}
 \includegraphics[width=0.5\textwidth]{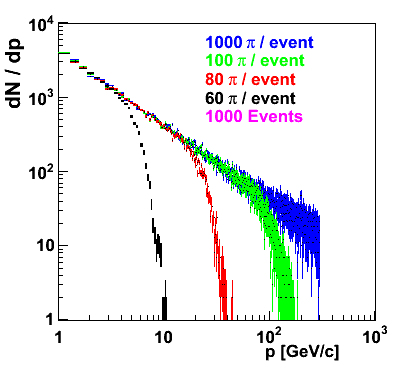}
\end{center}
\caption{Simulated $\pi^++\pi^-$ spactra from the model described in Sect.~\ref{sec:pccs}. The colour encoding distinguishes between the spectra of events with different multiplicities. Each histogram contains statistics of 1000 events. Through out the simulations, $a=1$ was used.}
 \label{fig:dNperdPa1}
\end{figure}

\begin{figure}
\begin{center}
 \includegraphics[width=0.5\textwidth]{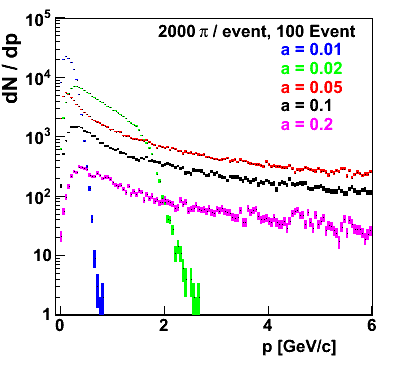}
\end{center}
\caption{Simulated $\pi^++\pi^-$ spactra from the model described in Sect.~\ref{sec:pccs}, shown in a log-lin plot. The colour encoding distinguishes between the spectra of events with different interaction measures, $a$. Each histogram contains statistics of 100 events with 2000 $\pi$-s per event.}
 \label{fig:dNperdPaLin}
\end{figure}

\begin{figure}
\begin{center}
 \includegraphics[width=0.5\textwidth]{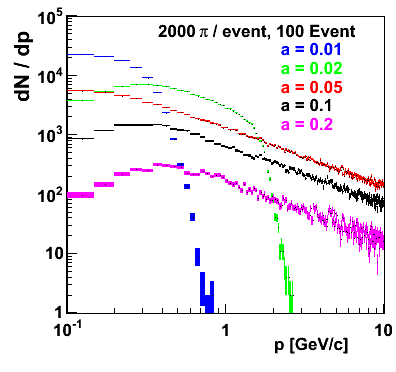}
\end{center}
\caption{Simulated $\pi^++\pi^-$ spactra from the model described in Sect.~\ref{sec:pccs}, shown in a log-log plot. The colour encoding distinguishes between the spectra of events with different interaction measures, $a$. Each histogram contains statistics of 100 events with 2000 $\pi$-s per event.}
 \label{fig:dNperdPaLog}
\end{figure}

The model described in Sect.~\ref{sec:pccs} converges to the TS distribution (Eq.~(\ref{onTS12}) in the case, when $L(\epsilon) = (1/a)\ln(1+a\epsilon)$)

\be
\frac{d\mathcal{P}_N}{dp} = A\,p^2\, (1+a\,\epsilon)^{-\beta/a} \, ,
\ee{res1}
when started from a homogeneously filled Fermi sphere. When resonance production (described in Sect.~\ref{sec:Rprod}) is added to the model as well, the results shown in Figs.~\ref{fig:dNperdM}-\ref{fig:dNperdPaLog} are obtained. As a result of the large interaction energy caused by the multi-particle interaction therms in Eq.~(\ref{onTS4}), the produced resonances may have large masses, as can be seen in Fig.~\ref{fig:dNperdM}. The mass distribution of the resonances, $dN_R/dM_R$ shows power-law behaviour for two orders of magnitude in the mass $M_R$. Low masses, $M_R < 2 m_{\pi}$ are not allowed throughout the formation.

When the resonances decay, their rest energy contributes the kinetic enrgies of the produced pions. This causes long power-law tails of the final state pion spectra shown in Figs.~\ref{fig:dNperdPa1}-\ref{fig:dNperdPaLog}. Fig.~\ref{fig:dNperdPa1} shows the dependence of the final state pion spectrum on the number of quarks (which is equal to the number of final state pions in the process $q\bar{q}\rightarrow R \rightarrow \pi^+\pi^-$) in the $QM$, when the interaction measure, $a=1$. Apparently, the spectra take the form of Eq.~(\ref{res1}) in the canonical region, where the one-particle energy is much smaller, than the total energy of the system, $\epsilon \ll E_N$. For larger energies, $\epsilon \gtrapprox \epsilon_0$, however the spectra have a cut. $\epsilon_0$ is proportional to the total energy, and grows like $\epsilon_0 \sim \bar{\epsilon}^{\,N}$.

Figs.~\ref{fig:dNperdPaLin} and \ref{fig:dNperdPaLog} show the dependence of the spectrum of events with fixed multiplicity ($N = 2000$) on the interaction measure $a$. Apparently, the spectra change from BG to TS, as the interaction measure grows from $a=0.01$ to $a=0.2$. The spectra with $a=0.01$ and $a=0.02$ are nearly BG distributions (apart from the cuts above $\epsilon_0$), while spectra with $a=0.05$, $a=0.1$ and $a=0.2$ are well-developed TS distributions (Eq.~(\ref{res1})).

\section{Conclusions}
\label{sec:con}
In this paper, we discussed hadronisation in a non-extensive quark matter with multiparticle interactions of the form of Eq.~(\ref{onTS4}). In Sect.~\ref{sec:onTS} we have shown that the one-particle distribution is the Tsallis distribution in a system, which is composed of independent and identically distributed particles, that are constrained on a constant energy hyper-surface in phase space by Eq.~(\ref{onTS4}). (We note, that this statement is valid only in the limit of a large number of particles, and if the one-particle energy is much smaller than the total energy of the system.) Furthermore, this statement holds for non-equi\-lib\-ri\-um systems as well (this, being its advantage over the approach based on the Maximum Entropy Principle (MEP)), however, the MEP may be used for non-independent particles too.

In Sect.~\ref{sec:pccs}, we have presented a hadronisation process, that respects the conservation of the total 3-momentum and non-extensive energy of the system. In the process, first resonances are formed, then decay into pion pairs. We have found that the resonances may have large masses, due to the large interaction energies in the quark matter, caused by the multiparticle interactions. This leads to the power-law tailed resonance mass distribution shown in Fig.~\ref{fig:dNperdM}. The resonance masses then contribute to the kinetic energies of the pions, resulting in long, power-law tailed pion spectra presented in Sect.~\ref{sec:res}. 

Since both our simulated and the measures pion spectra take the form of the Tsallis distribution, we may conclude, that the parameters of our model (mean energy per particle and interaction measure ($a$)) may be tuned so that our results and experimental data be in accordance with each other.

\section*{Acknowledgement}
\label{sec:ack}
This work was supported by the Hungarian OTKA grants PD73596, K68108 and the E\"otv\"os University. One of the authors (GGB) thanks the J\'anos Bolyai Research Scolarship of the Hungarian Academy of Sciences.

%

%
%


\begin{thebibliography}{99}

\bibitem{bib:tsinppNN1}
M. Shao, L. Yi, Z. Tang, H. Chen, C. Li, Z. Xu, J. Phys. G \textbf{37}, (2010) 085104

\bibitem{bib:tsinppNN2}
Z. Tang, Y. Xu, L. Ruan, G. Buren, F. Wang, Z. Xu, Phys. Rev. C \textbf{79}, (2009) 051901 (R)

\bibitem{bib:tsinppNN3}
D. D. Chinellato, J. Takahashi, I. Bediaga, J. Phys. G \textbf{37}, (2010), 094042

\bibitem{bib:tsinppNN4}
T. Wibig, J. Phys. G \textbf{37}, (2010) 115009

\bibitem{bib:tsinppNN5}
K. Urmossy, T. S. Biro, Phys. Lett. B \textbf{689}, (2010) 14-17

\bibitem{bib:tsinppNN6}
B De, G Sau, S. K. Biswas, S. Bhattacharyya, P. Guptaroy, J. Mod. Phys. A \textbf{25}, (2010) 1239-1251

\bibitem{bib:tsinppNN7}
J. Cleymans, G. Hamar, P. Levai, S. Wheaton, J. Phys. G \textbf{36}, (2009) 064 018

\bibitem{bib:tsinppNN8}
M. Biyajima, T. Mizoguchi, N. Nakajima, N. Suzuki, G. Wilk, Eur. Phys. J. C \textbf{48}, (2006) 597-603

\bibitem{bib:tsinppNN9}
G. G. Barnaf\"oldi, K. Urmossy, T. S. Biro, Proc. of Hot Quarks, (2010)

\bibitem{bib:tsinppNN9b}
G. G. Barnaf\"oldi, G. Kalm\'ar, K. Urmossy, T. S. Biro, Proc of Gribov '80 Workshop, (2010)

\bibitem{bib:tsinppNN10a} 
T. S. Biro, G. Purcsel, K. Urmossy, Eur. Phys. J. A \textbf{40}, (2009) 325-340

\bibitem{bib:tsinppNN10} 
T. S. Biro (ed.) \textit{et al.}, Eur. Phys. J. A \textbf{40}, (2009) 255-344

\bibitem{bib:tsinee1}
C. Beck, Physica A \textbf{286}, (2000) 164-180
\bibitem{bib:tsinee2}
I. Bediaga, E. M. F. Curado, J. M. Miranda, Physica A \textbf{286}, (2000) 156-163

\bibitem{bib:BEq1}
J. A. S. Lima, R. Silva, A. R. Plastino, Phys. Rev. Lett. \textbf{86}, (2001) 29-38
\bibitem{bib:BEq2}
G. Kaniadakis, Physica A \textbf{296}, (2001) 405; Phys. Rev. E \textbf{66}, (2002) 056125
\bibitem{bib:BEq3}
T. S. Biro, G. Purcsel, Phys. Rev. Lett. \textbf{95}, (2005) 162302

\bibitem{bib:noisy1}
T. S. Biro, A. Lakovac, Phys. Rev. Lett. \textbf{94}, (2005) 132302



\bibitem{bib:next1}
C. Tsallis, \textit{Introduction to Nonextensive Statistical Mechanics} (Springer, New York 2009)

\bibitem{bib:comp1}
T. S. Biro, Euro. Phys. Lett. \textbf{84}, (2008) 56003


\bibitem{bib:math1}
S. R. S. Varadhan, Asymptotic probability and differential equations, \textit{Comm. Pure Appl. Math.:} \textbf{19}, 261-286, 1966.

\bibitem{bib:math2}
Richard S. Ellis, Entropy, large deviations, and statistical mechanics, \textit{Springer-Verlag}, 1985.


\end{thebibliography}
\end{document}